\documentclass[11pt,a4paper]{article}
\usepackage[utf8]{inputenc}
\usepackage[T1]{fontenc}
\usepackage{lmodern}
\usepackage{microtype}
\usepackage[margin=2.5cm]{geometry}
\usepackage{setspace}
\usepackage{parskip}
\usepackage{graphicx}
\usepackage{booktabs}
\usepackage{enumitem}
\usepackage{hyperref}
\hypersetup{colorlinks=true,linkcolor=blue,citecolor=blue,urlcolor=blue}
\usepackage{titling}
\usepackage{fancyhdr}
\usepackage{caption}
\usepackage{amsmath,amssymb}
\usepackage[numbers,round]{natbib}

\newcommand*\mnras{MNRAS}
\newcommand*\nat{Nature}
\pretitle{\vspace*{-1cm}\begin{center}\LARGE\bfseries}
\posttitle{\par\end{center}\vskip 0.5em}
\preauthor{\begin{center}\large}
\postauthor{\end{center}}
\predate{\begin{center}\large}
\postdate{\end{center}}

\begin{document}

\begin{titlepage}
  \centering
  \vspace*{1cm}
  {\Huge\bfseries Expanding Horizons \\[6pt] \Large Transforming Astronomy in the 2040s \par}
  \vspace{1.5cm}

  {\LARGE \textbf{The multiple facets of millisecond pulsar binaries}\par}
  \vspace{1cm}

  \begin{tabular}{p{4.5cm}p{10cm}}
    \textbf{Scientific Categories:} & Time-domain, Neutron stars, Pulsars, Binaries \\
    \\
    \textbf{Submitting Authors:} 
    & Arianna Miraval Zanon \\
    & Affiliation: Agenzia Spaziale Italiana (IT) \\
    & Email: arianna.miraval@asi.it\\ \\
    & Giulia Illiano \\
    & Affiliation: INAF–OAB (IT)\\
    & Email: giulia.illiano@inaf.it \\
    \\
    \textbf{Contributing authors:}   
    & Filippo Ambrosino (INAF-OAR, IT)\\
    & Domitilla de~Martino (INAF-OACN, IT)\\
    & Maria~Cristina Baglio (INAF-OAB, IT)\\ 
    & Caterina~Ballocco (INAF-OAR, IT)\\
    & David Buckley (SAAO; University of Cape Town, ZA)\\
    & Francesco Coti~Zelati (ICE-CSIC, ES)\\
    & Melania Del Santo (INAF-IASF Palermo)\\
    & Paul J.~Groot (Radboud University, NL)\\
    & Riccardo La~Placa (INAF-OAR, IT)\\
    & Christian ~Malacaria (INAF-OAR, IT)\\
    & Alessio Marino (ICE-CSIC, ES) \\
    & Alessandro Papitto (INAF-OAR, IT) \\
    & Nanda Rea (ICE-CSIC, ES) \\
    & Andrea Sanna (University of Cagliari, IT) \\
    & Simone~Scaringi (Durham University, UK)\\
    & James~Turner (INAF-OAR, IT)\\
    & Alexandra~Veledina  (University of Turku, FI)\\
    & Luca Zampieri (INAF-OAPd, IT)\\
  \end{tabular}

\vspace{1cm}

\end{titlepage}


\section{Introduction and Background}

\label{sec:intro}
Millisecond pulsars (MSPs) are fast-spinning ($<$10~ms) and weakly magnetized ($10^{7-8}$~G) neutron stars (NSs), resulting from a Gyr-long accretion phase from a low-mass companion ($< 1$ M$_\odot$) [1]. The discovery of the accreting MSP (AMSP) SAX J1808.4-3658 [2] provided the first direct confirmation of this evolutionary link. AMSPs are transient systems that spend most of their time in quiescence (X-ray luminosity of $L_X \lesssim 10^{31-33} \, \mathrm{erg \, s^{-1}}$), but occasionally undergo accretion outbursts lasting days to months, with a strong X-ray brightening ($L_X \sim 10^{36-37} \, \mathrm{erg \, s^{-1}}$) and the presence of X-ray pulsations. These systems are unique laboratories to investigate matter and particle acceleration under extreme conditions that are unattainable on Earth. In these compact binaries with orbital periods typically shorter than one day, the interplay between the pulsar wind and the inflowing matter determines the MSP emission regime. If gravity dominates, the plasma is funneled onto the magnetic poles and accreted, producing X-ray pulsations; if the pulsar wind is strong enough to eject the matter, the system shines as a radio and $\gamma$-ray rotation-powered MSP. Observing the interaction between the NS magnetosphere and relativistic wind with inflowing matter at different mass loss rates is key to addressing fundamental questions: \textbf{Where do particles interact with the surrounding plasma to produce pulses at all energies? How are particles accelerated to relativistic energies by the pulsar wind?}

The connection between accretion-powered and rotation-powered pulsars has been strengthened by the discovery of transitional MSPs (tMSPs), which swing between these states on timescales of days [3,4,5]. tMSPs also exhibit an intermediate sub-luminous disk state in which accretion- and rotation-powered features coexist, including an accretion disk, intermediate X-ray luminosity, and $\gamma$-ray luminosity up to an order of magnitude higher than in typical rotation-powered MSPs [6 and refs. therein]. A signature of this state is erratic X-ray emission, with rapid switches between two intensity levels (``high'' and ``low'' modes) [7], and occasional flaring episodes. 
In the tMSP prototype PSR~J1023$+$0038 [3], coherent pulsations are detected simultaneously in X-ray, UV, and optical bands during high modes, vanishing in low modes [8,9,10,11,12]. Together with similar pulse profiles and a pulsed, polarized spectral energy distribution consistent with a single power law, this suggests a common emission mechanism. Neither rotation- nor accretion-powered mechanisms can individually explain all properties. This has led to a scenario in which high-mode multi-band pulsations arise from synchrotron radiation in a boundary region $\sim$100~km from the NS, where the relativistic wind interacts with inflowing matter [10,13; see also 14,15].

Beyond the tMSP PSR~J1023+0038, optical/UV pulsations have also been detected from the AMSP SAX J1808.4$-$3658 [16]. In the tMSP, optical and X-ray pulsations are in phase [14], while in the AMSP, they are in anti-phase. Moreover, PSR~J1023+0038 is observed in an inefficient accretion regime (i.e., sub-luminous disk state), while SAX~J1808.4$-$3658 undergoes active surface accretion during outburst. These differences indicate that optical pulsations in SAX~J1808.4$-$3658 are not directly tied to the X-ray ones, pointing to a separate, unidentified process and highlighting the diversity of MSP emission mechanisms. Optical pulsations have also been reported from the rotation-powered MSP PSR J2339$-$0533 [17].  

\section{Open Science Questions in the 2040s}
\label{sec:openquestions}
The discovery of three optical MSPs in different evolutionary states -- the accretion-powered, the rotation-powered, and the intermediate sub-luminous disk regimes [16,17,9] -- has raised new questions while offering the potential to address long-standing issues:
\begin{itemize}
    \item \textbf{Do all MSPs emit optical pulsations?} 
    With hundreds of MSPs known across all evolutionary stages, the few optical detections suggest that pulsations may be intrinsically rare or remain undetected due to instrumental limits and observational biases. Determining whether optical pulsations are universal or occur only under specific conditions has key implications for particle acceleration, magnetospheric emission, and the connection between optical and high-energy radiation.

    \item \textbf{How does the accretion disk increase the efficiency of converting spin-down power into pulsed optical emission?} For young, isolated rotation-powered pulsars, the conversion of spin-down power into pulsed optical emission ($\eta \simeq 10^{-5}-10^{-8}$; [18]) is well explained by magnetospheric synchrotron models. The efficiency of the optical rotation-powered MSP is comparable to that of the Crab pulsar, indicating that the same emission process likely operates in both classes [17]. In contrast, the optical tMSP and the AMSP are $\sim$50–100 times more efficient than predicted by rotation-powered models [10,17], implying that magnetospheric emission alone is insufficient. The presence of an accretion disk likely boosts particle acceleration [10,13], but with only one source per class, these conclusions remain tentative. A larger sample of optical MSPs across different accretion regimes is required to determine if the disk enhances efficiency or if the observed high efficiencies are peculiar to the few studied systems.
    
    \item \textbf{How do pulsar and disk interact? What diagnostics are needed?} 
    The intermediate sub-luminous disk state of tMSPs provides a rare opportunity to probe the interaction between accretion flow and pulsar wind when their energy densities are comparable. Yet, the disk structure's reaction in this state, especially during rapid X-ray mode switching, remains poorly understood. Discrete mass ejections have been invoked to explain the peculiar high-to-low mode switching [19], while compact jets or discrete ejections may power the flaring mode [15]. Switches between modes and flaring episodes can be investigated through Hydrogen and Helium emission lines in the UV/optical/nIR, tracing disk regions at different radii [20] . This requires high–time–resolution spectroscopy at moderate resolution to map the inner and outer disks and reveal the structural changes associated with the disappearance of pulsations during low modes. During flaring episodes, only optical pulsations have been detected without accompanying X-ray pulsations [11], making it crucial to trace the disk response during flares. Such diagnostics are essential to test disk thickening and the associated enhancement of free-electron injection into a compact jet.
    
    \item \textbf{How do the pulsar and companion wind interact, and how does the pulsar wind affect the companion's physical status and its evolution on different timescales, including transitions to/from disk state?} In rotation-powered MSPs, the interaction of the pulsar and companion winds produces an intrabinary shock (IBS),  observed primarily through orbitally modulated synchrotron X-ray emission. The IBS strongly interacts with the companion, where matter can be channeled along magnetic field lines. Optical emission features superimposed on the companion's absorption lines, often variable and correlated with irradiation-driven brightness changes, have been observed in an increasing number of systems and may originate in the IBS region [21]. These features still need to be explored through time-resolved spectroscopy to determine whether they trace variability in wind interactions that could lead to disk formation and the appearance of a tMSP. In particular, tracking the onset or inhibition of mass transfer, which signifies the system's transitional state, requires short (less than a day) reaction times and orbital-phase-resolved spectroscopy to monitor disk formation or disappearance in detail. The powerful pulsar wind in MSP binaries irradiates the companion star at varying levels, with the IBS contribution depending on the pulsar spin-down power and binary properties. Ablation of the companion is a key process in the evolutionary path of MSPs. While MSP binaries with low-mass companions ($<0.1\,$M$_{\odot}$) -- the black widows -- are likely at the final evolutionary stage, those with higher-mass, non-degenerate secondaries -- the redbacks, which include tMSPs -- are crucial systems for understanding the effects of irradiation on binary evolution. The increasing number of discovered systems, now exceeding a hundred, including candidates, with varying irradiation levels and IBS efficiencies, requires the determination of binary parameters and characterization of the companion star’s physical state through photometry and spectroscopy. While feasible for nearby systems, many newly discovered MSPs are more distant and faint, requiring large-aperture facilities and time-resolved observations.
\end{itemize}

\normalsize
\section{Technology and Data Handling Requirements}
\label{sec:tech}

\begin{itemize}
    \item {\bf High time resolution:} Discovering new optical MSPs and probing their rotational evolution requires an ultra-fast photometer with $\mu$s-level relative time resolution and $\leq$10~$\mu$s absolute timing accuracy. A 25-m class facility would improve photon-counting statistics by a factor of $\sim$60 over 4-m telescopes and, combined with semi-coherent search techniques, would enable the discovery of new MSPs in the optical band, even without precise estimates of the system’s orbital parameters [22].
    \item {\bf Medium-fast reaction and flexible scheduling:} some accretion phases last only days, requiring observations within hours of state transitions and a dynamic schedule for rapid, high-priority observations of transients.
    \item {\bf Setup:} the photometer should enable simultaneous observations of the target, a reference star, and the nearby sky ($\sim$1 arcmin), in white light and Sloan filters (u’, g’, r’, i’, z’), to study pulse variations and phase shifts across bands, with extension to the nIR to probe highly absorbed regions in the Galactic plane and bulge. Fast variability is expected in the nIR even though no current instrument reaches sub-millisecond resolution in this band.
    \item {\bf Medium spectral resolution} (R $\sim$ 5000--10000) is required to trace disk and IBS variability through line-profile changes, measure orbital periods via radial velocities, and construct spectral energy distributions to disentangle emission components.
    \item{\bf Polarimetry:} fast (ns–ms) optical and nIR photometry and polarimetry are crucial to study magnetized compact objects. Phase-resolved polarimetry ($\lesssim$0.03\% accuracy) constrains pulse emission geometry, while sub-second photometry probes particle acceleration, shocks, and magnetic reconnection at jet bases, linking outflows to inflows. These observations provide unique insight into the physics of NS systems.
\end{itemize}
All three instruments — the photometer, the spectrograph, and the polarimeter — are crucial for addressing the open questions outlined above. While each can operate independently, the ability to operate them simultaneously by 2040 would be even more advantageous.


\scriptsize
\textbf{References:} 
[1] Alpar, M.~A., Cheng, A.~F., Ruderman, M.~A., et al.\ 1982, \nat, 300, 5894, 728;
[2] Wijnands, R. \& van der Klis, M.\ 1998, \nat, 394, 6691, 344;
[3] Archibald A.~M., Stairs I.~H., Ransom S.~M., et al., 2009, Sci, 324, 1411;
[4] Papitto, A., Ferrigno, C., Bozzo, E., et al.\ 2013, \nat, 501, 7468, 517;
[5] Bassa, C.~G., Patruno, A., Hessels, J.~W.~T., et al.\ 2014, \mnras, 441, 2, 1825;
[6] Papitto A., de Martino D., 2022, ASSL, 465, 157;
[7] Linares M., 2014, ApJ, 795, 72;
[8] Archibald A.~M., Bogdanov S., Patruno A., et al., 2015, ApJ, 807, 62;
[9] Ambrosino, F., Papitto, A., Stella, L., et al.\ 2017, Nature Astronomy, 1, 854;
[10] Papitto A., Ambrosino F., Stella L., et al., 2019, ApJ, 882, 104;
[11] Jaodand A.~D., Hern{\'a}ndez Santisteban J.~V., Archibald A.~M., et al., 2021, arXiv, arXiv:2102.13145;
[12] Miraval Zanon A., Ambrosino F., Coti Zelati F., et al., 2022, A\&A, 660, A63;
[13] Veledina A., N{\"a}ttil{\"a} J., Beloborodov A.~M., 2019, ApJ, 884, 144;
[14] Illiano G., Papitto A., Ambrosino F., et al., 2023, A\&A, 669, A26;
[15] Baglio M.~C., Coti Zelati F., Di Marco A., et al., 2025, ApJL, 987, L19;
[16] Ambrosino F., Miraval Zanon A., Papitto A., et al., 2021, NatAs, 5, 552;
[17] Papitto A., Ambrosino F., Burgay M., et al., 2025, A\&A, 702, L21. doi;
[18] Mignani R.~P., 2011, AdSpR, 47, 1281;
[19] Baglio M.~C., Coti Zelati F., Campana S., et al., 2023, A\&A, 677, A30;
[20] Messa M.~M., D'Avanzo P., Coti Zelati F., et al., 2024, A\&A, 690, A344;
[21] Miraval Zanon A., D'Avanzo P., Ridolfi A., et al., 2021, A\&A, 649, A120;
[22] La Placa R., Papitto A., Illiano G., et al., 2025, arXiv, arXiv:2508.04873. doi:10.48550/arXiv.2508.04873, submitted to A\&A

\end{document}